# The transmission calculation by empirical numerical model and Monte Carlo simulation in high energy proton radiography of thick objects[*]

ZHENG Na (郑娜)     XU Hai-Bo (许海波)[1)]

Institute of Applied Physics and Computational Mathematics, Beijing 100094, China

**Abstract:**

An empirical numerical model that includes nuclear absorption, multiple Coulomb scattering, and energy loss is presented in this work to calculate the transmission of the thick objects in high energy proton radiography, and in this numerical model the angular distributions are treated as Gaussian in the laboratory frame. A Monte Carlo program based on Geant4 toolkit was developed using for the high energy proton radiography experiment simulation, and verifying the empirical numerical model. The two models are used to calculate the transmission fraction of the step-wedge of carbon and lead in proton radiography at 24GeV/c. The two models can reproduce the experiment data, and the differences are analyzed.

## 1 Introduction

High-energy proton radiography provides a new and quantitative technique for hydrotest experiments diagnoses [1]. Transmission radiographies with high spatial and resolution can be made if proton with tens of GeV energy illuminate a thick dynamic test object which is placed in the object plane of a point –to-point magnetic quadrupole imaging system [2].

The basic principle of proton radiography is the mechanism for proton-object interaction, including the energy loss, nuclear interaction, and multiple coulomb scattering. In proton radiography experiment, the attenuation of protons passing through an object is measured to analyze the material and the shape of the object. Because of the multiple coulomb scattering and the elastic scattering, the image will be blurred, and the definition will be reduced. In the design of the proton radiography experiment, the magnetic lens system and the collimators can be used to

---

[*] Supported by NSAF (11176001) and Science and Technology Developing Foundation of China Academy of Engineering Physics (2012A0202006)

[1)] E-mail：hbxu2002@yahoo.com.cn

control the proton beam, reduce the scattering exposure in the image plane, and sort the scattered beam in terms of how it has been scattered [3]. The density distribution and material composition of the object can be obtained from comparing the imaging radiographies from different collimators.

In the present work, an empirical forward model which considered the main essential physics in the proton radiography was developed to calculate the proton passing through thick objects. Meanwhile, a Monte Carlo approach based on Geant4 toolkit [4] was performed to simulate the proton radiography experiment for thick objects. The results of the two methods were compared.

## 2 The empirical numerical model

As mentioned above, the three most important effects on the protons as they go through an object are absorption, multiple Coulomb scattering, and energy loss. In the simplest form of the present model, a model for multiple Coulomb scattering (MCS) with a large-angle, small impact parameter cutoff due to the form factor of the nucleus plus strong- interaction attenuation are employed. And the scattering distributions are treated as Gaussian or sums of Gaussian in the laboratory frame scattering angle.

First, the angular distribution for a slab that is thin relative to the scattering channel is calculated, in order to derive expressions of the angular distributions for thick objects. When protons through a slab of material with thickness L (areal density) in the thin limit (L $\ll \lambda$ ),

$$\lambda = \frac{A}{N_A \sigma} \tag{1}$$

Where $\lambda$ is the mean free path for a particular channel, A is the atomic weight, $N_A$ is the Avogadro's number, and $\sigma$ is the cross section for the reaction channel.

there is only elastic scattering channel, and the angular distribution is

$$I_E(\theta) = \frac{\delta(\theta) + L / \lambda_E \ f_E(\theta)}{1 + L / \lambda_E} \tag{2}$$

where $\lambda_E$ is the mean free pass for elastic scattering, $\delta$ is the Dirac delta function, and $f_E(\theta)$ is the angular distribution for elastic scattering.

Second, the MCS is applied to the Eq. (2), the angular distribution is

$$I(\theta) = \frac{f_{MCS}(\theta) + L/\lambda_E \, (f_{MCS} * f_E)(\theta)}{1 + L/\lambda_E} \tag{3}$$

where $f_{MCS}(\theta)$ is the angular distribution for MCS, and $(f_{MCS} * f_E)(\theta)$ is the convolution of the elastic scattering angular distribution and the MCS angular distribution.

Third, the nuclear attenuation is considered, the angular distribution becomes

$$I(\theta) = \frac{f_{MCS}(\theta) + L/\lambda_E \, (f_{MCS} * f_E)(\theta)}{1 + L/\lambda_E} \, e^{-L/\lambda_A} \tag{4}$$

where $\lambda_A$ is the mean free pass for nuclear attenuation.

Finally, if the thickness L and the mean free pass $\lambda$ are comparable, the slab length L can be divided into N segments to satisfy the condition $(L/N \ll \lambda)$. Thus, the elastic scatter angular distribution of the slab can be approximated by the convolution of the elastic scattering angular distribution form one segment $L/N$ with itself N-1 times. And the angular distribution can be write as

$$I(\theta) = (\sum_{n=0} \frac{1}{n!} (\frac{L}{\lambda_E})^n (f_{MCS} * (f_E)^n)(\theta) \,) e^{-L/\lambda_A} \tag{5}$$

Thus, in the proton radiography, the angular collimator is placed in the Fourier plane of the magnification lens to control the scattering exposure in the image plane. Considering the effect angle cuts of the collimator, the transmission in the image plane of the proton passing the thick object is

$$T = 2\pi \int_0^{\phi_{cut}} I(\theta) d\theta \tag{6}$$

where the $\phi_{cut}$ is the angle of the collimator.

In the above empirical numerical model, the key is the scattering distributions calculation. In this work, the scattering distributions are treated as Gaussian in the laboratory frame, and the expression of Schiz et al. [5] and Baishev et al. [6] is used to the scattering distribution calculation.

## 3 The Monte Carlo simulation

In the present work, a program based on Geant4 toolkit was developed using for the proton radiography experiment simulation. The beam of 24 GeV/c with $10^{10}$ incident protons illumined the thick object for carbon and lead, and the setup of the simulation is shown in Fig. 1. First, the beam is prepared with a diffuser and monitor lens to meet optics requirements. Next the beam is controlled by the magnification imaging lens and collimator, and detected after it passes through the objected being radiographed.

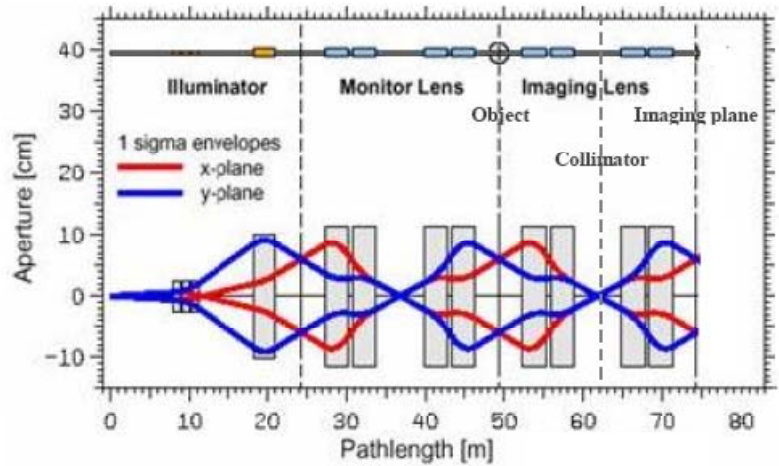

Fig. 1. The setup of the proton radiography used for the Monte Carlo simulation

In the simulation, the diffuser is a tantalum with 1.2 cm thickness. The magnification imaging lens consisted of 4 quadrupoles with 20 cm diameter and 120 cm long. The collimator is a cylinder of tungsten with 1.2 m long, and the collimator approximated multiple scattering angle acceptance cuts of 6.68 mrad. The main parameters of the setup for simulation are shown in Table 1.

Table 1. The main parameters of the setup for the present simulation.

| Configures of simulation | parameters |
|---|---|
| Proton momentum/GeV·c⁻¹ | 24 |
| Diffuser thickness/cm | 1.2 |
| Collimator length/m | 1.2 |
| Angular cut/mrad | 6.68 |
| Quadroupole aperture/mm | 120 |
| Quadroupole gradient/T·m⁻¹ | 8 |
| Quadroupole length/m | 2 |
| Dirft length/m | 3.4 |
| Chromatic aberration coefficient/m | 40.4 |
| Field of view/mm | 80 |

In the Monte Carlo simulation, the hadronic shower models provided in Geant4 are used to compute the interaction of the proton and the object, and the Runge-Kutta method is used to track the proton trajectory in the magnetic field.

## 4 Comparison of the two models with experimental data

The results of the empirical numerical model and the Monte Carlo simulation were compared with the transmission fractions extracted from step- wedge data for various elements from EA955 experiment at Brookhaven National Laboratory [7]. In this paper, the objects are step-wedge for carbon and Lead with the areal density form 0 to 26 cm. The comparisons of the transmission fraction of the two models and the experimental data are shown in Fig. 2.

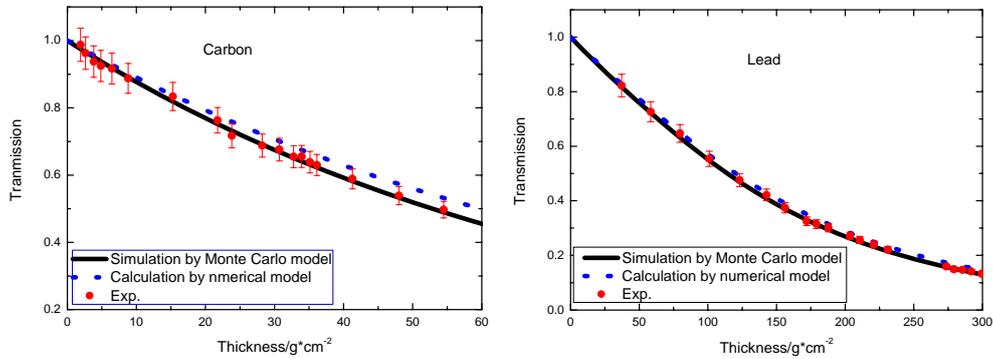

Fig. 2. Comparison of the calculation of the Monte Carlo model and the numerical model and the experimental data for the carbon and lead material.

The figures show that the results for transmission versus areal density from the two models are in good agreement with the experiment EA955 at the AGS proton accelerator at Brookhaven National Laboratory.

The figures also show that the transmission calculated by the numerical model is a little larger than the Monte Carlo simulation and the experimental data. There are two aspects that induced the differences. On the one hand, the quasi-elastic scattering process in the proton interaction physics is not included in the numerical model, and on the other hand, In the Monte Carlo simulation and the proton radiography experiment, there are the transportation through the quadrupole magnets, while the numerical model just imposed a cut on the particles based on the angle that a given particle has as it left the object, this assumption is different with the experiment.

## 5 Conclusions

The present empirical numerical model which includes the essential physics of all of the elements of the radiographic chain, and the Monte Carlo model that based on Geant4 toolkit are developed and employed to calculate the transmission of the high energy proton in passing through a thick slab of material. The two models are agreed with experimental step wedge proton transmission data from the experimental data. The calculations also show there are few differences between the numerical mode and the Monte Carlo model, which come from the contribution of the quasi-elastic scattering process and the angle-cut assumption. Thus, the Monte Carlo model can be used to predict the experiment of the proton radiography, and the empirical numerical model can be used to verify the calculation of Monte Carlo model. In the next work, the quasi-elastic scattering process will be considered in the numerical model to improve the calculation.

# 厚客体质子照相穿透率的数值计算和蒙卡模拟

**ZHENG Na (郑娜)    XU Hai-Bo（许海波）**

**摘要：**质子照相技术是流体动力学试验的一种先进诊断方法。本工作考虑弹性散射，多次库仑散射，核反应等物理过程对质子束流的衰减和角分布的影响，假定散射角分布均呈高斯形，用数值方法计算了 24GeV/c 的质子在 0-26cm 的碳、铅客体中的穿透率。同时基于 Geant4，开发了质子照相蒙特卡罗用户程序，对 24GeV/c 的质子在 0-26cm 的碳、铅客体中的传输进行了模拟。并将两种方法的计算结果与布鲁克海文国家实验室的测量进行了比较，结果与实验符合。

**关键词：**质子照相，角分布，散射，穿透，Geant4